\documentclass[12pt]{article}
\pdfoutput=1

\usepackage{graphicx,array}
\usepackage{color}
\usepackage{latexsym}
\usepackage{amsthm}
\usepackage{amsmath}
\usepackage{amssymb}
\usepackage[numbers,sort&compress]{natbib}
\usepackage{bm}
\usepackage{slashed} 
\usepackage{mathrsfs}
\usepackage{hyperref} 
\hypersetup{
    colorlinks=true,       
    linkcolor=red,          
    citecolor=blue,        
    filecolor=magenta,      
    urlcolor=blue           
}
\usepackage[all]{hypcap} 

\usepackage[font={small}]{caption}

\pdfoutput=1
\usepackage{graphicx}
\usepackage{epstopdf}
\usepackage{dcolumn}
\usepackage{bm}
\usepackage{hyperref}
\usepackage{color}
\usepackage{amsmath}
\usepackage{mathrsfs}
\usepackage{cancel}
\usepackage{xpatch}

\newcommand{\be}{\begin{equation}}
\newcommand{\ee}{\end{equation}}
\newcommand{\Op}{\mathcal{O}}

\begin{document}
\thispagestyle{empty}

\begin{flushright}
\end{flushright}
\bigskip 
\begin{center}
{\Large \bf $B$-decay anomalies  in Pati-Salam SU(4)} \\
\vspace{1cm}
{\large Riccardo Barbieri$^a$ and Andrea Tesi$^b$} \\
\vspace{.4cm}
{$^a$\it \small Scuola Normale Superiore, Piazza dei Cavalieri 7, 56126 Pisa, Italy and INFN, Pisa, Italy} \\
{$^b$\it \small INFN sezione di Firenze, Via G. Sansone 1, I-50019 Sesto Fiorentino, Italy }
\vspace{1cm}
\end{center}

\begin{abstract}\noindent
Attempts to incorporate in a coherent picture the $B$-decay anomalies presumably observed in $b\rightarrow c$ and $b\rightarrow s$ semi-leptonic decays have to face the absence of signals in other related experiments, both at low and at high energies. By extending and making more precise the content of Ref.~\cite{Barbieri:2016las}, we describe one such attempt based on the Pati-Salam SU(4) group, that unifies colour and the $B$-$L$ charge, in the context of a new strongly interacting sector, equally responsible for producing a pseudo-Goldstone Higgs boson. \\
\end{abstract}


\vfill
\noindent\line(1,0){188}
{\scriptsize{ \\ \texttt{$^a$ \href{mailto:riccardo.barbieri@sns.it}{riccardo.barbieri@sns.it}\\ $^b$ \href{mailto:andrea.tesi@fi.infn.it}{andrea.tesi@fi.infn.it}}}}

\newpage

\section{Introduction}

While no single experiment is precise enough to allow the claim of a discovered anomaly in $B$-decays, a combination of different experimental results in $b\rightarrow c$ and $b\rightarrow s$ semi-leptonic decays make altogether a case for an observed deviation from the Standard Model (SM) in flavour physics. We refer in particular to the presumed anomalies in $R_{D^{(*)}}^{\tau\ell}, \ell=\mu, e$ \cite{Lees:2013uzd,Aaij:2015yra,Hirose:2016wfn,ARV} and $R_{K^{(*)}}^{\mu e}$ \cite{Aaij:2014ora,Aaij:2017vbb}. To be able to speak of an overall significant case, however, one needs at least to offer a coherent description of both anomalies, capable, at the same time, to account for the absence of signals in several potentially correlated experiments. This is the aim of this work, extending and making more precise the content of Ref.~\cite{Barbieri:2016las}. The main feature of this attempt is the Pati-Salam SU(4) group \cite{Pati:1974yy}, which, in different ways and at different levels of depth, has also been considered as a relevant ingredient in several other recent works on the subject~\cite{Diaz:2017lit,DiLuzio:2017vat,Assad:2017iib,Calibbi:2017qbu,Bordone:2017bld}.

The experimental results/bounds that have to be kept in mind, among others, in conjunction with an explanation of $R_{D^{(*)}}^{\tau \ell}$, and $R_{K^{(*)}}^{\mu e}$ are:
\begin{itemize}
\item The couplings of $ b_L, \tau_L, \nu_\tau$ to the $Z$  and of the $\tau$ to the $W$, none of which deviates from the SM at relative $10^{-3}$ level;
\item The absence of deviations from the SM expectation in $\Delta B=2$ transitions;
\item The bound $\mathcal{B}(K_L\rightarrow \mu e) < 4.7\cdot 10^{-12}$;
\item The lack of signals  in direct production at LHC so far of possible mediators of the extra interactions that appear to be needed to make sense of the $20\div 30 \%$ deviation in $R_{D^{(*)}}^{\tau l}$,  described, in the SM, by a tree level $W$-exchange. 

\end{itemize}

As well known a key feature that may hide new flavour phenomena in whatever Beyond the Standard Model (BSM) physics is the U(3)$^5$ symmetry of the gauge sector of the SM, at the origin of so called Minimal Flavour Violation \cite{Chivukula:1987py,Hall:1990ac,DAmbrosio:2002vsn}.
The exponent in U(3)$^5$ refers to the number of irreducible representations of the SM gauge group under which one generation of matter transforms. With Yukawa couplings switched on, an approximate symmetry subgroup of U(3)$^5$ is U(2)$^5$ which acts on the first two generations of matter as doublets and on the third generation as singlets \cite{Barbieri:2011ci,Barbieri:2012uh,Barbieri:2012tu}. U(2)$^5$ is an observed approximate symmetry both of the masses of the charged fermions as of the mixing angles in the quark sector. Suppose now that a leptoquark be the mediator responsible for the $B$-anomalies, singlet under U(2)$^5$. In absence of U(2)$^5$ breaking the leptoquark can only couple to the third generation. This hypothesis provides a natural first-order explanation for the different size of the charged current versus the neutral current effects, about equally deviating from the tree-level versus loop-level SM amplitudes respectively: $b\rightarrow c\tau\nu$ only involves a single second generation particle, whereas $b\rightarrow s\mu\mu$ has three light generation fermions \cite{Barbieri:2015yvd}. As we shall see this feature can be extended, always by U(2)$^5$, to other vector mediators.

Let us insist on a leptoquark and, to make the game more constrained, let us suppose that it be a vector. The relatively large deviation from a charged current process, tree level in the SM, requires it not to be  too heavy. Can one make sense of such a vector in an acceptable theoretical framework? The leptoquark $\hat{V}_\mu$ in the adjoint of Pati-Salam SU(4) comes obviously to mind, transforming as $(3,1)_{2/3}$ of the SM gauge group. Two features, however, have to be taken into account, as already mentioned. The mediator cannot even be too light to have escaped detection at LHC so far. In turn, since in a low energy exchange what counts is the ratio of coupling versus mass, its coupling must be sizeable, at least $3\div 4$. Furthermore the process $K_L\rightarrow \mu e$ is mediated at tree level by $\hat{V}_\mu$-exchange at an unacceptable rate, if $(s, \mu)$ and $(d, e)$  form quartets of SU(4) with equal coupling as the third generation $(b, \tau)$. A possible way out from this last difficulty consists in invoking the presence of heavy Dirac quartets of SU(4) to which the standard quarks and leptons are suitably mixed. These features - a leptoquark with a strongish coupling and heavy Dirac fermions - lead us to consider both of them as composite particles of a new strong dynamics, assumed to exist, with SU(4) as a global symmetry group. As a further natural step, one can view this same new strong dynamics as responsible for producing, by spontaneous symmetry breaking,  a light pseudo-Goldstone Boson Higgs \cite{Agashe:2004rs, Giudice:2007fh}. The overall global symmetry that we consider is therefore SU(4)$\times$SO(5)$\times $U(1)$_X$ broken down to SU(4)$\times$SO(4)$\times$U(1)$_X$.

The paper is organized as follows. In Section \ref{def} we define the model making essential use of the CCWZ formalism to incorporate the non-linearly realized SU(4)$\times$SO(4)$\times$U(1)$_X$ symmetry. This brings in further vectors than $\hat V_\mu$. In Section \ref{heavyvec} we write down the couplings of the heavy vectors to the standard fermions. In Section \ref{observed} we show how the observed anomalies are accounted for. In Section \ref{delta_g_WZ} we show how the model can explain the absence of deviations in the couplings of $ b_L, \tau_L, \nu_\tau$ to the $Z$  and of the $\tau$ to the $W$ at about $10^{-3}$ level. In Section \ref{other_flavour} we discuss other especially constrained flavour observables. In Section \ref{DirectLHC} we give a preliminary discussion of the ongoing search at LHC of some of the vector states, with special emphasis on the single production of heavy gluons.
Finally in Section \ref{summary} we summarise the overall picture.

\section{The composite model defined}
\label{def}
The model is defined by the global symmetries, the representation of the resonances, and the coupling of the SM to the composite sector. The symmetry breaking pattern $G/H$ is defined by
\be
G=\mathrm{SU(4)}\times \mathrm{SO(5)}\times \mathrm{U(1)}_X,\quad H=\mathrm{SU(4)}\times \mathrm{SO(4)}\times \mathrm{U(1)}_X
\ee
and the model is characterized by the presence of vector and Dirac fermion fields in the following representations of $H$
\be
\mathcal{G}_\mu^A = (15,1)_0,\quad \rho_\mu^a = (1,6)_0,\quad X_\mu = (1,1)_0,\quad \Psi_\pm= (4,4)_{\pm \frac12}, \quad \chi_\pm= (4,1)_{\pm \frac12}.
\ee
The SM symmetry is introduced by an explicit breaking of $H$: the QCD sector is an SU(3)$_c$ subgroup of SU(4), the electroweak SU(2) is inside SO(4).
The hypercharge is $Y=T_{15}+T^3_R+X$, where $T_{15}=\mathrm{diag}(1,1,1,-3)/\sqrt{24}$ is a diagonal generator of SU(4), and $T_{3R}$ is one generator of the SU(2)$_R$ subgroup in SO(4).

In formulas, the embedding is such that the decomposition of the above resonances under SU(3)$_c\times$SU(2)$_L\times$U(1)$_Y$ is
\begin{eqnarray}\label{dec}
\mathcal{G}_\mu^A &\to& \hat{G}_{(8,1)_0} \oplus \big[\hat V_{(3,1)_{\frac23}} + c.c.\big] \oplus \hat B_{(1,1)_0}\,,\\
\rho_\mu^a &\to& \rho^L_{(1,3)_0} \oplus \rho^{R3}_{(1,1)_0} \oplus \rho^{R\pm}_{(1,1)_\pm}\,,\\
\Psi_+ &\to& Q_{(3,2)_{\frac{1}{6}}} \oplus X_{(3,2)_{\frac{7}{6}}} \oplus  L_{ (1,2)_{-\frac{1}{2}}} \oplus L^x_{ (1,2)_{1}}\,,\\
\Psi_- &\to& Q'_{(3,2)_{\frac{1}{6}}} \oplus X'_{(3,2)_{\frac{-5}{6}}} \oplus  L'_{ (1,2)_{-\frac{1}{2}}} \oplus L^y_{ (1,2)_{-1}}\,,\\
\chi_+ &\to& U_{(3,1)_{\frac{2}{3}}} \oplus N_{(1,1)_{0}}\,,\\
\chi_- &\to& D_{(3,1)_{\frac{-1}{3}}} \oplus E_{(1,1)_{-1}} \,,
\end{eqnarray}
where we have specified the name and the representation of the resonances that will appear later on. This already shows which are the new fields that can mix with the SM ones.
The mass splitting between the various representation is governed by the explicit breaking of the symmetry.

\subsection{The CCWZ Lagrangian}
We introduce the phenomenological Lagrangian of the model according to the Callan Coleman Wess Zumino (CCWZ) formalism \cite{Callan:1969sn}, which allows us to discuss non-linearly realized symmetries\footnote{This approach is at variance with the one of Ref.~\cite{Barbieri:2016las}, where the Higgs boson is treated as a generic composite particle rather than a specific pseudo-Goldstone boson. This difference is important in the structure of flavour and in the  electroweak constraints on $W$ and $Z$ couplings. 
A particularly useful reference for the use of the CCWZ formalism throughout this paper is \cite{Panico:2015jxa}.  See also  \cite{Grojean:2013qca,Matsedonskyi:2014iha}}. The main ingredient is the GB field  of SO(5)/SO(4), $U=\exp(i\sqrt{2}h_i \hat{T}^i/f)$), where $\hat{T}^i$ are the broken generators in the fundamental representation of SO(5), while the unbroken ones are $T^a$.  From this field we can construct the $d$ and $e$ symbols of the CCWZ formalism,
\be
i U^\dag D_\mu U= e^a_\mu T^a + d_\mu^i \hat{T}^i = (A_\mu^a  + \bar e^a_\mu )T^a + d_\mu^i \hat{T}^i,
\ee
where  $D_\mu$ is the covariant derivative of the (elementary) SM. On the contrary, SU(4) is an exact global symmetry of the composite sector before the gauging of the SU(3)$_c$ and U(1)$_Y$,  so that the $e$ and $d$ symbols are trivial: the former corresponds to the elementary gluon and hypercharge fields, while the latter vanishes.

The lagrangian of the model is the sum of three pieces
\be
\mathscr{L}=\mathscr{L}_{\rm ele}+\mathscr{L}_{\rm comp}+\mathscr{L}_{\rm flavour},
\label{Ltot}
\ee
that are defined in the following way.

The elementary sector, $\mathscr{L}_{\rm ele}$, is given by the SM lagrangian without the Yukawa and Higgs sectors
\be
\mathscr{L}_{\rm ele}=-\frac{1}{4g_2^2} W_{\mu\nu}^2-\frac{1}{4g_1^2} B_{\mu\nu}^2-\frac{1}{4g_3^2} G_{\mu\nu}^2 + \mathrm{fermionic\ kinetic\ terms}
\ee

We describe the composite sector with CCWZ formalism, including at the same time vector and fermionic resonances, all assumed  lighter than the cutoff. 
\begin{eqnarray}\label{Lcomp}
\mathscr{L}_{\rm comp}&=& \frac{f^2}{4}d_\mu^i d^{\mu,i} \nonumber\\
&-& \frac{\rho_{\mu\nu}^a \rho^{\mu\nu,a}}{4g_\rho^2}+\frac{m_\rho^2}{2g_\rho^2}(\rho_\mu^a-e_\mu^a)^2 - \frac{\mathcal{G}_{\mu\nu}^A \mathcal{G}^{\mu\nu,A}}{4g_G^2}+\frac{m_\mathcal{G}^2}{2g_G^2}(\mathcal{G}_\mu^A-G_\mu^A)^2 - \frac{X_{\mu\nu} X^{\mu\nu}}{4g_X^2}+\frac{m_X^2}{2g_X^2}(X_\mu-B_\mu)^2\nonumber\\
&+& i \bar\Psi_\pm \gamma^\mu (D_\mu -i \bar{e}_\mu^a T^a)\Psi_\pm - m_\psi \bar\Psi_\pm \Psi_\pm\nonumber\\ 
&+& c_G (\mathcal{G}_\mu^A - G_\mu^A) \bar\Psi _\pm\gamma^\mu \bar T^A \Psi_\pm+ c_\rho (\rho_\mu^a - e^a_\mu)\bar\Psi_\pm\gamma^\mu T^a \Psi_\pm+ \frac{c_X}{2} (X_\mu - B_\mu) \bar\Psi_\pm\gamma^\mu  \Psi_\pm\nonumber\\
&+& i \bar\chi_\pm \gamma^\mu D_\mu \chi_\pm - m_\chi \bar\chi_\pm \chi_\pm+ c_G (\mathcal{G}_\mu - G_\mu)^A \bar\chi_\pm\gamma^\mu \bar T^A \chi_\pm+ \frac{c_X}{2} (X_\mu - B_\mu) \bar\chi_\pm\gamma^\mu  \chi_\pm\nonumber\\
&+& i c d_\mu^i \bar\Psi_\pm^i \gamma^\mu \chi_\pm + \mathrm{h.c.}
\end{eqnarray}
where $T^a$ are the generators of SO(4), and $\bar{T}^A$ of SU(4).
Notice that the term with one $d$ symbol is only allowed when a fourplet and a singlet of SO(4) are present in the Lagrangian (see also \cite{Grojean:2013qca,Matsedonskyi:2014iha} for the inclusion of such term).
These formulas are presented in the formal limit where the entire group $G$ is gauged: in order to get the correct results one can simply set to zero the components of $A_\mu^a$ and $G_\mu^A$ that are not gauged (for example only color and hypercharge fields ($G^a, B$) are non zero among the  $G^A$). 
A flavour index running from 1 to 3 is  left understood in the fermion fields.

Finally, the flavour sector is given by the interaction between the elementary and composite sectors. The interactions break the global symmetries of the theory and will induce the SM Yukawa terms,
\begin{eqnarray}\label{Lflavour}
\mathscr{L}_{\rm flavour} &=&  \lambda_q f \, \bar{q}_L U \hat \Psi_{+}   + \lambda_u f \, \bar{u}_R U^\dag \hat \Psi_{+} +\lambda'_q f \, \bar{q}_L U \hat\Psi_{-} +\lambda_d f \, \bar{d}_R U^\dag \hat\Psi_{-} \nonumber \\
&+& \lambda_l f \, \bar{l}_L U \hat \Psi_{+}   + \lambda_l f \, \bar{l}_L U \hat \Psi_{-}  + \lambda_e f \, \bar{e}_R U^\dag \hat \Psi_{-}   + \mathrm{h.c.},
\end{eqnarray}
where $\hat{\Psi}=(\Psi,\chi)$ is a fiveplet. For brevity of notation we leave understood a relative coefficient between $\Psi$ and $\chi$ in every term, in general flavour dependent.

 By working in the background where $U$ is equal to the identity we can resolve the leading order mixing between the elementary and the heavy fields. The terms in \eqref{Lflavour} that play an important role in the following are the ones originating from $ \lambda_q f \, \bar{q}_L U \Psi_{+} $ and $ \lambda_l f \, \bar{l}_L U \Psi_{+} $. They induce a mixing between $q_L$ and $Q_L$ in $\Psi_{+L}$ and between  $l_L$ and $L_L$ also in $\Psi_{+L}$, respectively described by angles with sines given by
\be
s_q\equiv \sin\theta_q = \frac{\lambda_q f}{\sqrt{m_\Psi^2 +\lambda_q^2 f^2}},\quad\quad
s_l\equiv \sin\theta_l = \frac{\lambda_l f}{\sqrt{m_\Psi^2 +\lambda_l^2 f^2}}
\ee
 All other mixing angles are taken sufficiently small that they can be neglected in the couplings of the heavy vectors to the standard fermions.  With these extra angles exactly vanishing, only the up quarks get mass $m_u=s_q\lambda_u V/\sqrt{2}$ and, in the limit of exact $U(2)^n$ (see below), only the top mass survives, $m_t=s_{q3}\lambda_{u3} V/\sqrt{2}$. Had we introduced right-handed neutrinos, $\nu_R$, and the interaction $\lambda_\nu \bar{\nu}_RU^+ \hat \Psi_{+}$, neutrinos too would have received a Dirac mass $m_\nu = s_l\lambda_\nu V/\sqrt{2}$. A large Majorana mass $M \nu_R \nu_R$ gives then rise to the usual see-saw mechanism. For the present purposes the $\nu_L$ can be taken massless and coincident with the  eigenstates of the standard weak interactions. Note that the down and the charged lepton masses arise from mixing with the states in $\hat \Psi_{-}$. This is essential to keep under control the corrections to the couplings of $b_L, \tau_L$ to the $Z$, as discussed below.
%

\subsection{Flavour}

The elementary sector, $\mathscr{L}_{\rm ele}$, has a $U(3)^5$ flavour symmetry distinguishing the five different irreducible representations of the SM gauge group. Similarly $\mathscr{L}_{\rm comp}$ is assumed to respect a $U(3)^{\hat{\Psi}}\times U(3)^{\hat{\chi}}$ flavour symmetry. All the breaking of flavour is confined to $\mathscr{L}_{\rm flavour}$, which, as suggested by the observed flavour parameters (masses and mixings),  is assumed to possess a weakly broken symmetry
\begin{equation}
U(2)^n \equiv U(2)^{\hat{\Psi}}\times U(2)^{\hat{\chi}}\times U(2)^q\times U(2)^l\times U(2)^u\times U(2)^d\times U(2)^e.
\end{equation}
In absence of $U(2)^n$-breaking only the third generation of elementary fermions   mixes  with the composite fermions. Via this mixing, therefore, 
the heavy vectors only couple to the third generation of elementary fermions. As shown below, this may explain: i) why anomalies show up, at least so far, only in $B$-decays, and ii) why, within $B$-decays, $b\rightarrow c\tau\nu$ and $b\rightarrow s\mu\mu$ show deviations from the SM of similar size, in spite of being, in the SM, a tree level and a loop process respectively.

\subsection{Mass eigenstates in the vector sector}

With the exception of the leptoquark $\hat{V}_\mu$ and the charged $\rho^{R\pm}_\mu$, all other mass eigenstates in the vector sector are admixtures of the elementary and composite vectors. For example the properly normalized massless and massive gluons are
\be
 g_\mu^a = \frac{ g_G G_\mu^a+ g_3\mathcal{G}_\mu^a}{\sqrt{g_G^2+g_3^2}},\quad\quad
  \hat{G}_\mu^a = \frac{g_G\mathcal{G}_\mu^a -  g_3 G_\mu^a  }{\sqrt{g_G^2+g_3^2}}
  \ee
  and similarly for the other vectors. 
  The massive vectors whose composite component is in SU(4) or SU(2)$\times$SU(2) or U(1)$_X$ have masses $m_G, m_\rho, m_X$ up to corrections of order $(g_3/g_G)^2$ or $(g_2/g_\rho)^2, (g_1/g_\rho)^2$ or $(g_1/g_X)^2$ respectively. At the same time $g_3, g_2, g_1$ can be identified with the standard strong, weak and hypercharge couplings up to similar corrections.

\section{Couplings of the heavy vectors to the light fermions}
\label{heavyvec}

There are two sources of couplings of the heavy vectors with the light fermions: one due to mixing in the fermion sector and one due to mixing in the vector sector.

\subsection{By mixings in the fermion sector}

As said, $Q_L$ and $L_L$ are the only heavy fermions significantly mixed with  $q_L$ and $l_L$ by $s_q$ and $s_l$ respectively. 
With reference to the Lagrangian \eqref{Lcomp}, setting 
\be
\hat{g}_G = g_G c_G,\quad \hat{g}_\rho = g_\rho c_\rho,\quad \hat{g}_X = g_X c_X,
\ee
the couplings of the heavy composite vectors to these composite fermions are
\begin{eqnarray}\label{non-universal-couplings}
\mathscr{L}_{\rm int}^G &=& \frac{\hat{g}_G}{\sqrt{2}} (\hat{V}_\mu (\bar{Q}_L\gamma_\mu L_L) + \mathrm{h.c.})+
\frac{\hat{g}_G}{2} \hat{G}^a_\mu (\bar{Q}_L\gamma_\mu \lambda^a Q_L) +
\frac{\hat{g}_G}{2\sqrt{6}} \hat{B}_\mu (\bar{Q}_L\gamma_\mu Q_L - 3 \bar{L}_L\gamma_\mu L_L)\,,\\
\mathscr{L}_{\rm int}^\rho &=& \frac{\hat{g}_\rho}{2} \hat{\rho}_{L\mu}^\alpha (\bar{Q}_L\gamma_\mu \sigma^\alpha Q_L + \bar{L}_L\gamma_\mu \sigma^\alpha L_L) -
\frac{\hat{g}_\rho}{2} \hat{\rho}_{R\mu}^3 (\bar{Q}_L\gamma_\mu Q_L + \bar{L}_L\gamma_\mu L_L)\,,\\
\mathscr{L}_{\rm int}^X &=& 
\frac{\hat{g}_X}{2} \hat{X}_{\mu} (\bar{Q}_L\gamma_\mu Q_L + \bar{L}_L\gamma_\mu L_L) \,.
\end{eqnarray}
In the interaction basis, after integrating out the heavy fermions,  one gets from these equations the interactions of the heavy bosons with $q_L$ and $l_L$ by
\begin{equation}\label{rotation-fermions}
Q_L\rightarrow s_q q_L,\quad\quad L_L\rightarrow s_l l_L
\end{equation}
Without loss of generality we work in the basis where $s_q$ and $s_l$ are diagonal in flavour space
 \begin{equation}
 s_{q} = (s_{q1}, s_{q2}, s_{q3}),\quad\quad s_{l} = (s_{l1}, s_{l2}, s_{l3})
 \end{equation}
with, by weak $U(2)^n$-breaking, $s_3\gg s_{2,1}$.
\subsection{By mixings in the vector sector}

A further source of coupling of the heavy vectors with the light fermions, this time flavour independent, comes from the mixing among the composite and the elementary vectors. These interactions are
\be
\mathscr{L}_{\rm int}^{\rm mix}=
-\frac{g_3^2}{g_G}\hat{G}_\mu^a J_\mu^{3a}
-\frac{g_2^2}{g_\rho}\hat{\rho}_\mu^{La }J_\mu^{2a}
-g_1^2 (\frac{1}{g_G}\hat{B}_\mu+ \frac{1}{g_\rho}\hat{\rho}_\mu^{3R}+\frac{1}{g_X}\hat{X}_\mu) J_\mu^{1a}
\ee
where $J_\mu^{(3,2,1)}$ are the SM currents associated with the SU(3)$_c$, SU(2)$_L$ and U(1)$_Y$ groups respectively. In the relevant parameter space, these couplings give subdominant or at most comparable contributions to some flavour changing observables (see Section \ref{other_flavour}). On the contrary, for $s_{q1}< s_{q2}$ they are the leading couplings of the first generation of quarks to the heavy gluons, relevant to its direct production (see Section \ref{DirectLHC}).

\section{Observed anomalies}
\label{observed}

We call $U,D,E$ the unitary matrices that diagonalise the Yukawa couplings on the left side
\begin{equation}\label{rotazioni}
y_{u} = U^\dag y^{\rm diag}_u U_R\,,\quad y_{u} = D^\dag y^{\rm diag}_d D_R\,,\quad y_{e} = E^\dag y^{\rm diag}_e E_R\,.
\end{equation}
By $U(2)^n$ these matrices have small  elements that connect the third to the first and second generations. The unitary transformations on the right do not play any role in the following discussion.

The effects of the exchanges of the composite bosons are described at tree level  by the following effective Lagrangians. In writing down these Lagrangians we keep the dominant terms by expanding in the flavour parameters. The diagonal elements of $U, D, E$ are approximated to unity, thus fixing a phase convention.
For consistency with the constraints coming from $\Delta B=2$ transitions (see Section \ref{other_flavour}), we shall in fact approximate the full $D$ matrix to unity. In this limit the CKM matrix is given by $V_{ij}\approx U_{ij}$.

\subsection{$B$ to $D^{(*)}$ semileptonic decays}
The leading new physics contribution to $B\to D^{(*)}\tau \nu$ transitions arises from the exchange of charged resonances:
the leptoquark $\hat V$ in the $t$-channel and the charged $\hat{\rho}^{L\pm}$ in the $s$-channel,
\begin{equation}\label{bctaunu}
\mathscr{L}_{b\rightarrow c\tau\nu}=- (\frac{\hat{g}_G^2}{2m_G^2} + \frac{\hat{g}_\rho^2}{2m_\rho^2}) s^2_{q3} s^2_{l3} V_{cb}^* (\bar{c}_L\gamma_\mu b_L) (\bar{\tau}_L\gamma_\mu \nu_L).
\end{equation}
We can write the above term in a gauge invariant way, by anticipating the use of the effective Lagrangian
\begin{equation}
\mathscr{L}^{\rm eff}_{4f} = -\frac{1}{V^2} \big(C_3 \mathcal{O}^{3}_{lq} + C_1 \mathcal{O}^{2}_{lq}\big)\label{L4f}
\end{equation}
where $V=246\,\mathrm{GeV}$ and the two (fermion current)$\times$(fermion current) operators are given by
\begin{equation}
\mathcal{O}^3_{lq}=  (\bar{q}_{L3}\gamma_\mu \sigma^\alpha q_{L3})
(\bar{l}_{L3}\gamma_\mu \sigma^\alpha l_{L3}),\quad
\mathcal{O}^1_{lq}= (\bar{q}_{L3}\gamma_\mu  q_{L3})
(\bar{l}_{L3}\gamma_\mu  l_{L3}).
\end{equation} 
The Wilson coefficients of these operators can be computed at the scale where the vector resonances are integrated out by making use of eq.s~\eqref{non-universal-couplings}-\eqref{rotation-fermions}. Although only $\mathcal{O}^3_{lq}$ contributes to $B\to D^{(*)}$, we here write both coefficients for later convenience
\begin{equation}
C_3 = \frac{V^2}{4} (\frac{\hat{g}_G^2}{m_G^2} + \frac{\hat{g}_\rho^2}{m_\rho^2} )s_{l3}^2s_{q3}^2, \quad
C_1 = \frac{V^2}{4} (\frac{\hat{g}_G^2}{2m_G^2} + \frac{\hat{g}_\rho^2}{m_\rho^2}+ \frac{\hat{g}_X^2}{m_X^2} )s_{l3}^2s_{q3}^2.
\label{C3C1}
\end{equation}
It is worth noticing that the vectors in SO(4) contribute in an equal amount to $C_1$ and $C_3$.

For the decays into muons one gets
\begin{equation}\label{bcln}
\mathscr{L}_{b\rightarrow c\mu\nu}=-  s^2_{q3} s^2_{l3} V_{cb}^* [\frac{\hat{g}_G^2}{2m_G^2} (|E_{\mu 3}|^2 +\frac{E_{\mu 3}}{V_{cb}^*}\frac{s_{q2}s_{l2}}{s_{q3}s_{l3}})+ \frac{\hat{g}_\rho^2}{2m_\rho^2} (|E_{\mu 3}|^2 +(\frac{s_{l2}}{s_{l3}})^2)]
(\bar{c}_L\gamma_\mu b_L) (\bar{\mu}_L\gamma_\mu \nu_L)
\end{equation}
and similarly for $b\rightarrow c e\nu$ with $E_{\mu 3}$ replaced by $E_{e 3}$ and $s_2$ by $s_1$. The richer structure of the coefficient of the effective operator in eq.~\eqref{bcln} as compared to \eqref{bctaunu} can be understood when decomposed into two sources: a contribution originates from eq.~\eqref{bctaunu} using the rotation $E$ of eq.~\eqref{rotazioni} and the other arises by the coupling to second generation leptons.

\medskip
For small enough  $U(2)^n$-breaking parameters, $E_{\mu 3}, E_{e 3}$ and $s_2, s_1$, as discussed below, the charged current anomaly can be expressed in terms of $C_3$ as \cite{Buttazzo:2017ixm,Amhis:2016xyh}
\begin{equation}
R_D^{(*)} = 1 + 2C_3 = 1.237\pm 0.053.
\label{RD}
\end{equation}
Similarly, deviations from Lepton Flavour Universality (LFU) in the first two generations are constrained by \cite{Buttazzo:2017ixm,PDG}
\begin{equation}
R^{\mu/e}_{b\rightarrow c} = 1.000\pm 0.021,
\end{equation}
which, however, does not pose significant constraints on  $E_{\mu 3}, E_{e 3}$ and $s_2, s_1$.


\subsection{$B\rightarrow K^{(*)} \ell\ell$ semileptonic decays}
The deviations observed in $B\rightarrow K^{(*)} \mu\mu$ originate, in this framework, from the exchange of several mediators. However, because of the constraints from $\Delta B_s=2$ observables (discussed in section \ref{other_flavour}), we are led to consider the unitary matrix $D$ very close to unity. This fact, together with $s_{q,l 2}\ll s_{q,l 3}$, gives
\begin{equation}
\mathscr{L}_{b\to s \mu\mu}=
-\frac{\hat{g}^2_G}{2m_G^2}s_{q2}s_{l2}s_{q3}s_{l3} E_{\mu3} (\bar{s}_L\gamma_\mu b_L)(\bar{\mu}_L\gamma_\mu \mu_L),
\end{equation}
with the leptoquark exchange as the only responsible for the anomaly in $R_K^{(*)}$.

Therefore, using the result of \cite{Capdevila:2017bsm} for the relevant Wilson coefficient
\begin{equation}
\Delta C^\mu_9=-\Delta C^\mu_{10} = -\frac{2\pi}{\alpha V_{tb}V_{ts}^*}\frac{C_3}{1+ (\frac{\hat{g}_\rho m_G}{\hat{g}_G m_\rho})^2}s_{q2}s_{l2}s_{q3}s_{l3} E_{\mu 3}=-0.61\pm0.12
\end{equation}
i.e., using (\ref{RD})
\begin{equation}
\frac{s_{q2}s_{l2}}{s_{q3}s_{l3}}\frac{E_{\mu3}}{V_{ts}}\sim
5\cdot 10^{-3} 
\label{RK}
\end{equation}
with $\hat{g}_G/m_G$ and $ \hat{g}_\rho/m_\rho$ taken comparable.

Similarly, also the rate for $B\rightarrow K^{(*)} \tau\tau$ has a small deviation from the SM. Furthermore the corrections to  $B\rightarrow K^{(*)} \nu\nu$ vanish  at tree level and are sufficiently suppressed at loop level even with a cutoff from the composite dynamics $\Lambda = \mathcal{O}(10)$ TeV.

\section{Electroweak constraints on $W$ and $Z$ couplings}
\label{delta_g_WZ}
Renormalisation group running down to the weak scale of the Lagrangian (\ref{L4f}) corrects the couplings of the $W,Z$ to the third generation fermions \cite{Feruglio:2017rjo}, which, at low energies, are affected by the following two operators
\begin{equation}
\mathscr{L}^{\rm eff}_{Hl} = \frac{1}{V^2}(\bar{C}_{3l} \mathcal{O}^3_{Hl}
+ \bar{C}_{1l} \mathcal{O}^1_{Hl})
\label{C13l}
\end{equation}
where the (Higgs current)$\times$(fermion current) operators are defined in Table \ref{operators}. For quarks the situation is completely analogous. Notice that the bar over the Wilson coefficients indicates that they are evaluated at the electroweak scale, differently from the unbarred ones of \eqref{L4f}, which are instead generated at the scale of the resonances.

It is worth emphasizing that eq.~\eqref{L4f} does not exhaust all the possible effective operators that can induce, via running and mixing, distortions in the electro-weak boson couplings. All the operators of Table \ref{operators} modify  these couplings, with the same obvious inclusion of the `tree-level' contribution of eq.~\eqref{C13l} at the high scale. Such contribution is  calculated explicitly in Appendix A. 
\begin{table}[t]\small
$$\begin{array}{c|c|c}
\hbox{name}& \hbox{structure} & \hbox{coefficient} \\
\hline
\Op_{Hl}^{1}  &  i H^\dag \overleftrightarrow{D_\mu} H  (\bar l_{L3} \gamma^\mu l_{L3}) & C_{1l}\\
\Op_{Hl}^{3}  &  i H^\dag \sigma^a \overleftrightarrow{D_\mu} H (\bar l_{L3} \gamma^\mu \sigma^a l_{L3})& C_{3l}\\
\Op_{Hq}^{1}  &  i H^\dag \overleftrightarrow{D_\mu} H  (\bar q_{L3} \gamma^\mu q_{L3})& C_{1q}\\
\Op_{Hq}^{3}  &  i H^\dag \sigma^a \overleftrightarrow{D_\mu} H (\bar q_{L3} \gamma^\mu \sigma^a q_{L3})& C_{3q}\\ \hline
\Op_{ql}^{1}  &  (\bar q_{L3} \gamma^\mu q_{L3})(\bar l_{L3} \gamma^\mu l_{L3})& - C_{1}\\
\Op_{ql}^{3}  &  (\bar q_{L3} \gamma^\mu \sigma^a q_{L3})(\bar l_{L3} \gamma^\mu \sigma^a l_{L3})& - C_{3}\\
\Op_{ll}  & (\bar l_{L3} \gamma^\mu l_{L3})^2& C_{ll} \\
\Op_{qq}^{1}  & (\bar q_{L3} \gamma^\mu q_{L3})^2& C_{1qq} \\
\Op_{qq}^{3}  & (\bar q_{L3} \gamma^\mu \sigma^a q_{L3})^2 & C_{3qq}
\end{array}$$
\caption{\label{operators} The 9 operators relevant for the electro-weak fit. They define an effective lagrangian $V^2 \mathscr{L}=\sum_i C_i \Op_i$.}
\end{table}

Specifically, the four-fermion operators of Table \ref{operators} renormalize the $\mathcal{O}_{Hl}, \mathcal{O}_{Hq}$ operators due to SM gauge and Yukawa interactions. Therefore, upon integration of  the RG equations given in Appendix \ref{running}, this implies that the low-energy parameters of eq.~\eqref{C13l} are functions of the coefficients in Table \ref{operators}.
In turn, this can be used  to directly compute the modifications of the $W/Z$ couplings to third generation fermions
\begin{eqnarray}\tiny
\delta g^Z_\tau &=& -\frac{C_{3l}+C_{1l}}{2} -0.040 C_1 + 0.040 C_{1l} + 0.034 C_3 + 0.021 C_{3l} + 
 0.00035 C_{ll}\,,\\
\delta g^Z_\nu &=& \frac{C_{3l}-C_{1l}}{2} -0.040 C_1 + 0.040 C_{1l} - 0.034 C_3 - 0.021 C_{3l} - 
 0.0041 C_{ll}\,,\\
\bigg|\frac{g^W_\tau}{g^W_l}\bigg| &=& 1+C_{3l}-0.067 C_3 - 0.043 C_{3l} - 0.0045 C_{ll}\,,\\
\label{dgZb}
\delta g^Z_b &=& -\frac{C_{3q}+C_{1q}}{2}+ 0.00062 C_1 + 0.056 C_{1q} + 0.082 C_{1qq} - 0.0021 C_3 + 
 0.097 C_{3q} - 0.012 C_{3qq}\,.\nonumber \\
\end{eqnarray}
The above shifts are normalized to the SM in the following way
\be
\frac{g}{c_W} Z_\mu (g_{f_L}^{\rm SM}+ \delta g^Z_{f_L}) J^\mu_f  + \big[\frac{g}{\sqrt{2}}  W_\mu ( g^W_\tau \bar{\nu}_L\gamma^\mu \tau_L + g^W_l \bar{\nu_l}_L\gamma^\mu l_L) + \mathrm{h.c.}\big]\,.
\ee
The experimental constraints on these deviations are \cite{ALEPH:2005ab,Erler:2017ozu}
\be\label{data-ew}
\delta g^Z_{\tau_L}=  -0.0002\pm0.0006\,,\quad\quad  \delta g^Z_{\nu_\tau}=  -0.0015\pm 0.0013\, \quad\quad |g^W_\tau/g^W_l | = 1.00097\pm 0.00098.
\ee
One can then use these results to constrain combinations of Wilson coefficients of the operators listed in Table \ref{operators}. In principle one would expect to determine three independent linear combinations. However, the fact that only the  two operators in eq. (\ref{C13l}) can affect at low energy the lepton couplings reflects itself in the relation
\be
\delta g_\nu^Z - \delta g^Z_{\tau_L} = |g^W_\tau/g^W_l|-1,
\ee
no matter what the input values for the Wilson coefficients are at the high scale.\footnote{Including corrections that are quadratic in these deviations, one can in principle gain sensitivity to more than two operators, but given the present constraints this is irrelevant.} 

\begin{figure}[t]
\begin{center}
\includegraphics[width=0.45\textwidth,height=0.4\textwidth]{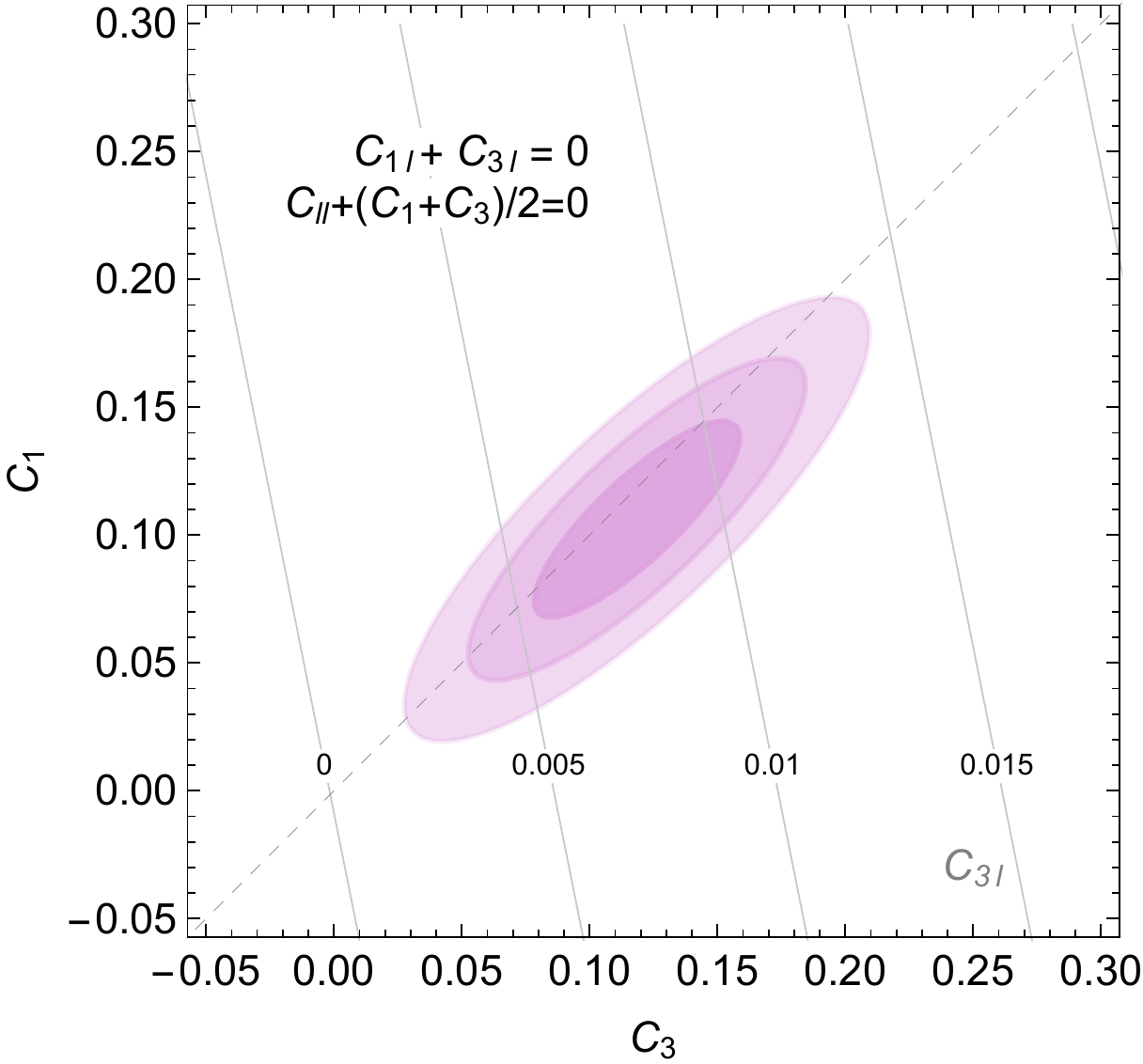}~
\includegraphics[width=0.45\textwidth,height=0.4\textwidth]{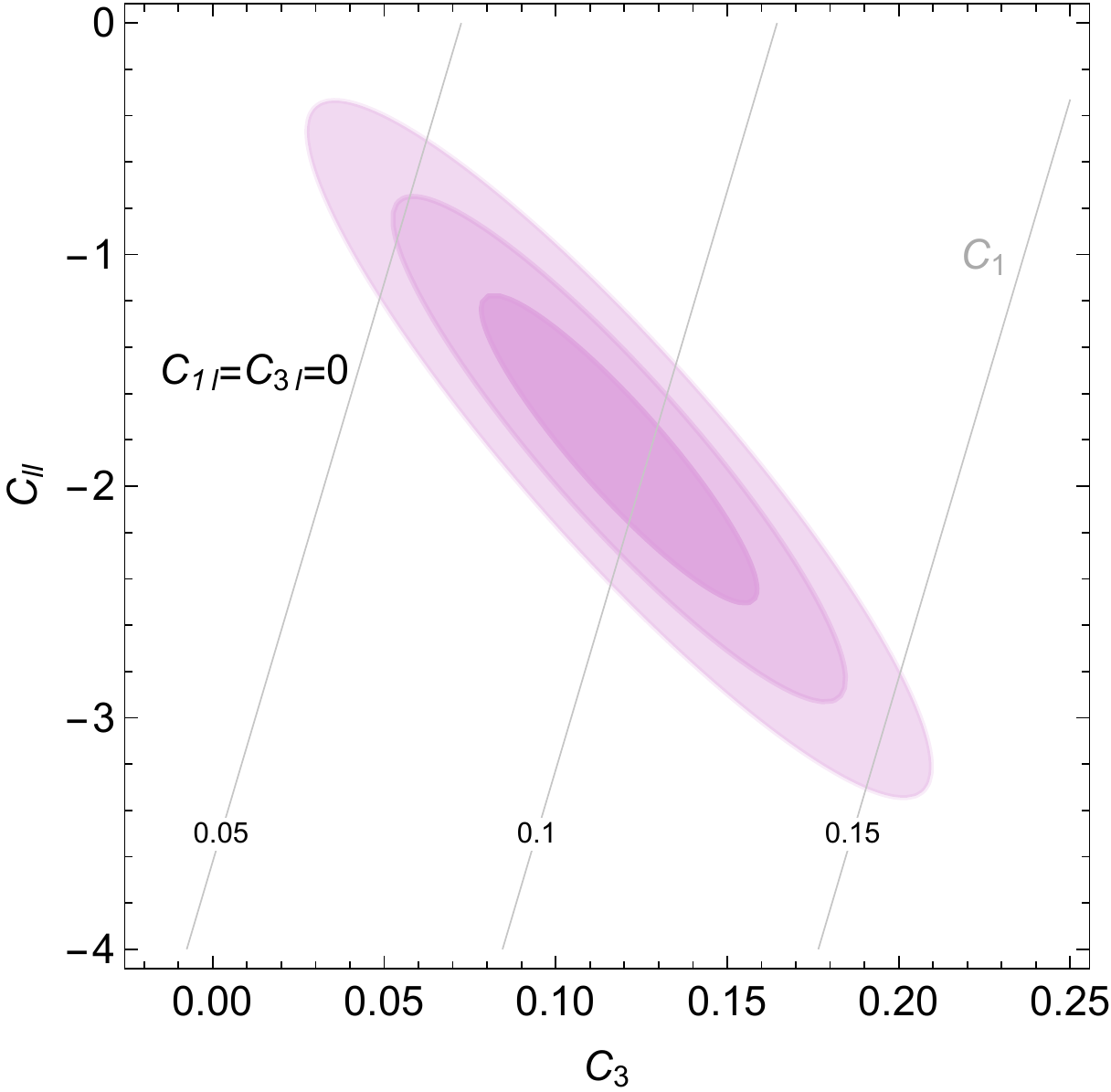}
\caption{\label{FIT} Fit of $R_D^{(*)}$, eq. (\ref{RD}), and of the corrections to $W/Z$ couplings of eq.~\eqref{data-ew}. Left panel: fit to $C_3, C_1$ and $C_{3l}=-C_{1l}$ with $C_{ll}=-(C_1+C_3)/2$. The full isolines are for $C_{3l}$. The dotted line is for $C_1=C_3$. Right panel: fit to $C_3, C_1$ and $C_{3l}=C_{1l}=0$. The isolines are for $C_1$.}
\end{center}\end{figure}

\medskip
In our model, thanks to its global symmetries and parities \cite{Agashe:2006at}, we have $C_{3l}=-C_{1l}$ as well as $C_{3q}=-C_{1q}$ from the composite sector. Furthermore, in analogy with eq. (\ref{C3C1}), the exchange of the neutral vectors gives a contribution to $C_{ll}$ that is a linear combination of $C_{1,3}$,
\be
C_{ll}=- \frac{V^2}{2}(\frac{3\hat{g}_G^2}{8m_G^2} + \frac{\hat{g}_\rho^2}{2m_\rho^2}+
\frac{\hat{g}_X^2}{4m_X^2} )s_{l3}^4 = -\frac{1}{2} (\frac{s_{l3}}{s_{q3}})^2(C_1 +C_3).
\ee

Based on the experimental values \eqref{data-ew} and on $R_D^{(*)}$ as in (\ref{RD}), we show in the left panel of Fig.~\ref{FIT} a fit of all these quantities in terms of $C_{3l}=-C_{1l}$ and $C_3, C_1$, having fixed $C_{ll}=-1/2(C_1+C_3)$ or $s_{l3}=s_{q3}$. The success of this fit depends crucially on the choice of the representations under the global group of the composite fermions. As shown in Appendix \ref{AA},
 $C_{3l}$ and $C_{1l}$ get contributions both from the $e_\mu$ and the $d_\mu$ terms in $\mathscr{L}_{\rm comp}$, eq. (\ref{Lcomp}), which must be of opposite sign and can partially cancel among each other so that $C_{3l}=-C_{1l}\approx 0.1 (C_3, C_1)$ as required  by the fit. Note that a non vanishing $C_{3l}=-C_{1l}$ is needed. A fit with $C_{3l}=-C_{1l}=0$ would require an anomalously large value of $C_{ll}$, as shown in the right panel of Fig.~\ref{FIT}.  We do not include in the fit the coupling $\delta g_{b_L}^Z = (3.3\pm 1.6) 10^{-3}$ \cite{Ciuchini:2014dea}, due to the presence in eq.~\eqref{dgZb} of the coefficient $C_{3q}=-C_{1q}$ which is otherwise unconstrained.

%
%

The fit only determines two ratios between the three low energy parameters $\hat{g}_G/m_G, \hat{g}_\rho/m_\rho$ and $\hat{g}_X/m_X$. However, the close correlation $C_1\approx C_3$ indicates $\hat{g}_X/m_X\approx 1/\sqrt{2}\hat{g}_G/m_G$. Furthermore, as already noticed, the fact that the leptoquark exchange is the only one responsible for the anomaly in $R_K$, suggests to take $ \hat{g}_\rho/m_\rho \lesssim \hat{g}_G/m_G$. With these inputs,  we take in the following the reference value 
\be
\hat{g}_G s_{q3} s_{l3} = 2 \frac {m_G}{\rm TeV},
\label{preferred}
\ee
knowing that some variations are possible.

\subsection{Other precision constraints and related tuning}
In the present model other precision tests  can in principle help to constrain the framework. From the non-linearity of the pseudo-Goldstone boson Higgs we expect to generate at the high scale the operator terms
\be
\frac{c_H}{2f^2} (\partial_\mu  |H|^2)^2 + (\frac{c_y y_f}{f^2} \bar f_L H f_R |H|^2 + \mathrm{h.c.}).
\ee
They both modify the Higgs couplings to vectors and fermions, and $c_H$ renormalizes the operators that contribute to the $\epsilon_{1,3}$ parameters \cite{Barbieri:2007bh}. In this model we have $c_H=1$ and $c_y=1$, leading to
\be\label{higgs}
\frac{g_{hVV}}{g_{hVV}^{\rm SM}}= 1- \frac{1}{2}\frac{V^2}{f^2},\quad \frac{g_{hff}}{g_{hff}^{\rm SM}}= 1- \frac{3}{2}\frac{V^2}{f^2},
\ee
and
\be\label{st}
\hat{T}\approx \Delta \epsilon_1 = -\frac{3\alpha}{8\pi c_W^2}\frac{V^2}{f^2}\log\frac{M}{m_h}.,\quad\hat{S}\approx \Delta \epsilon_3 =  \frac{\alpha}{24\pi s_W^2} \frac{V^2}{f^2} \log \frac{M}{m_h}\,.
\ee
As well known, other corrections to the electroweak parameters depend on  UV physics \cite{Grojean:2013qca}, which makes it  conceivable that the present constraints on eq.s~\eqref{higgs}-\eqref{st} are satisfied for $f\gtrsim 700\ \mathrm{GeV}$, i.e. a minimal amount of tuning (computed as $\sim f^2/V^2$ to 5$\div$10\%).

Notice that $V^2/f^2$ is not directly constrained by the previous fit, although it enters, among other parameters, the expressions for $C_{3l,1l}$ , as computed in App.~\ref{AA}. However, one can argue that, barring O(1) factors, we need $\hat{g}_G/m_G \gtrsim 2 /f$, in order to comply with the above constraints, and a mild cancellation in $C_{3l,1l}$ in order to minimize the bound on $f$.

\section{Other low energy flavour observables}
\label{other_flavour}
The presence of non universal couplings of the vector resonances to quarks and leptons generates four fermion operators, in addition to the ones already discussed in section \ref{observed}, which contributes to other  flavour transitions. A list of relevant constraints, together with the corresponding bound on the leading effective operator, is given in Table \ref{others}. For $\Delta B_s =2$ and $\Delta C =2$ we use the stronger bounds, depending on the phase of the corresponding coefficient \cite{Isidori:2013ez}. For the remaining coefficients we use  \cite{PDG}
\be
\mathcal{B}(K_L\rightarrow \mu e) < 4.7\times 10^{-12},\quad
\mathcal{B}(\tau\rightarrow 3 \mu ) < 2.1\times 10^{-8},\quad
\mathcal{B}(\tau\rightarrow \mu \gamma) < 4.4\times 10^{-8}.
\ee

\begin{table}[h]\small
$$\begin{array}{c|c|c}
\hbox{Observables}& \hbox{Operators} & \hbox{Bound}\\
\hline  \hline  
\Delta B_s =2  & C_{bs} (\bar{s}_L\gamma_\mu b_L)^2 & C_{bs}\mathrm{TeV^{2}} < 1.7\times 10^{-5} \\
\Delta C =2  &  C_{uc}(\bar{c}_L\gamma_\mu u_L)^2 & C_{uc} \mathrm{TeV^{2}}< 1.1\times 10^{-7}  \\
\mathcal{B}(K_L\to \mu e)  & C_{K\mu e} (\bar{s}_L\gamma_\mu d_L) (\bar{\mu}_L\gamma_\mu e_L) &C_{K\mu e} \mathrm{TeV^{2}}<  10^{-5}  \\
\mathcal{B}(\tau \to 3\mu)  & C_{\tau3\mu}(\bar{\mu}_L\gamma_\mu \tau_L)(\bar{\mu}_L\gamma_\mu \mu_L) &C_{\tau3\mu} \mathrm{TeV^{2}}< 3\times 10^{-3} \\
\mathcal{B}(\tau\to \mu \gamma)  & C_{\tau\mu\gamma} m_\tau e F_{\mu\nu}(\bar{\mu}_L\sigma_{\mu \nu}\tau_R)  & C_{\tau\mu\gamma} \mathrm{TeV^2}< 0.9\times 10^{-3} \\
\end{array}$$
\caption{\label{others} Relevant constraints from low energy flavor data.}
\end{table}

\subsection{$\Delta B_s=2$}
For the $\Delta B_s=2$ transitions the effective Lagrangian is
\begin{equation}
\mathscr{L}_{\Delta B_s=2}=- \frac{1}{2}(\frac{3\hat{g}_G^2}{8m_G^2} + \frac{\hat{g}_\rho^2}{2m_\rho^2}+
\frac{\hat{g}_X^2}{4m_X^2} ) D^2_{s3} s^4_{q3} (\bar{s}_L\gamma_\mu b_L)^2
\end{equation}
Using the central value of eq. (\ref{preferred}), together with $\hat{g}_X/m_X\approx 1/\sqrt{2}\hat{g}_G/m_G$ and small $ \hat{g}_\rho/m_\rho$,  
consistency with the bound in Table \ref{others} requires
\begin{equation}
\frac{s_{q3}}{s_{l3}} D_{s3}\lesssim 4\cdot 10^{-3}
\end{equation}
against $V_{ts}\approx U_{t2}+D_{s3}= 4\cdot 10^{-2}$. This motivates to take $D=1$ in the following, for which we offer no explanation. Note  that $D=1$ significantly suppresses $\Delta B=2$ transitions also at (quadratically divergent) loop level.

\subsection{$\Delta C=2$}

In the case of $D-\bar{D}$ mixing, the relevant effective Lagrangian is
\begin{equation}
\mathscr{L}_{\Delta C=2}=- \frac{1}{2}(\frac{3\hat{g}_G^2}{8m_G^2} + \frac{\hat{g}_\rho^2}{2m_\rho^2}+
\frac{\hat{g}_X^2}{4m_X^2} )(s_{q3}^2 V_{cb} V_{ub}^* + s_{q2}^2 V_{us}^*V_{cs})^2 (\bar{c}_L\gamma_\mu u_L)^2\,.
\end{equation}
Consistency with Table \ref{others} requires
\begin{equation}
(\frac{s_{q3}}{s_{l3}} V_{cb} V_{ub}^* + \frac{s_{q2}^2}{s_{q3}s_{l3}} V_{us}^*V_{cs}) \lesssim 3\cdot 10^{-4}\,.
\label{boundDeltaC=2}
\end{equation}

\subsection{$K_L\rightarrow \mu e$}
Insisting on $D=1$ also in the 1-2 sector, it is
\begin{equation}
\mathscr{L}_{s\rightarrow d \mu e}=-  s_{q2} s_{l2} s_{q1} s_{l1} \frac{\hat{g}_G^2}{2m_G^2} 
(\bar{s}_L\gamma_\mu d_L) (\bar{\mu}_L\gamma_\mu e_L).
\end{equation}
Consistency with Table \ref{others} requires
\begin{equation}
\frac{\sqrt{s_{q2} s_{l2} s_{q1} s_{l1}}}{s_{q3}s_{l3}}\lesssim 2\cdot 10^{-3}\,.
\end{equation}

\subsection{$\tau\rightarrow 3 \mu$}
LFV $\tau$ decays proceed both via heavy vector exchanges and via $Z$-exchange with modified $Z\tau\mu$ coupling. 
The relevant Lagrangian from 
 tree level neutral vector exchanges is
\begin{equation}
\mathscr{L}_{\tau\rightarrow 3 \mu,\rm tree} = - \frac{1}{2}(\frac{3\hat{g}_G^2}{8m_G^2} + \frac{\hat{g}_\rho^2}{2m_\rho^2}+
\frac{\hat{g}_X^2}{4m_X^2} )  s^2_{l3} E_{\mu3} (s^2_{l2} + s_{l3}^2|E_{\mu3}|^2)(\bar{\mu}_L\gamma_\mu \tau_L)(\bar{\mu}_L\gamma_\mu \mu_L)\,.
\end{equation}
With this effective Lagrangian in isolation, consistency with Table \ref{others} requires
\begin{equation}
 E_{\mu3} (\frac{s^2_{l2}}{s_{l3}^2} + |E_{\mu3}|^2) \lesssim 3\cdot 10^{-3}\,.
 \label{boundtau3mu}
\end{equation}
Considering also the $Z$-exchange contribution in isolation  and  taking $\delta g^Z_\tau\approx 10^{-3}$, the contraint from $\tau\rightarrow 3 \mu$ is satisfied for a weaker bound  $E_{23} \lesssim 0.3$.

\subsection{$\tau\rightarrow \mu\gamma$}

From one loop radiative corrections one gets
\begin{equation}
\mathscr{L}_{\tau\rightarrow \mu\gamma}=  \frac{1}{16\pi^2}E_{\mu3} m_\tau (A_G s^2_{q3}s^2_{l3} \frac{\hat{g}_G^2}{m_G^2} + A_\rho s^4_{l3}  \frac{\hat{g}_\rho^2}{m_\rho^2})
\, e F_{\mu\nu}(\bar{\mu}_L\sigma_{\mu \nu}\tau_R) \,,
\end{equation}
where $A_G, A_\rho$ are order 1 coefficients. Consistency with Table \ref{others} requires
\begin{equation}
(A_G +(\frac{s_{l3}}{s_{q3}})^2A_\rho) E_{\mu 3}\lesssim 0.4\times 10^{-1}\,.
\label{boundtaumugamma}
\end{equation}

\section{Signals at the LHC}
\label{DirectLHC}

\begin{figure}
\includegraphics[width=0.45\textwidth]{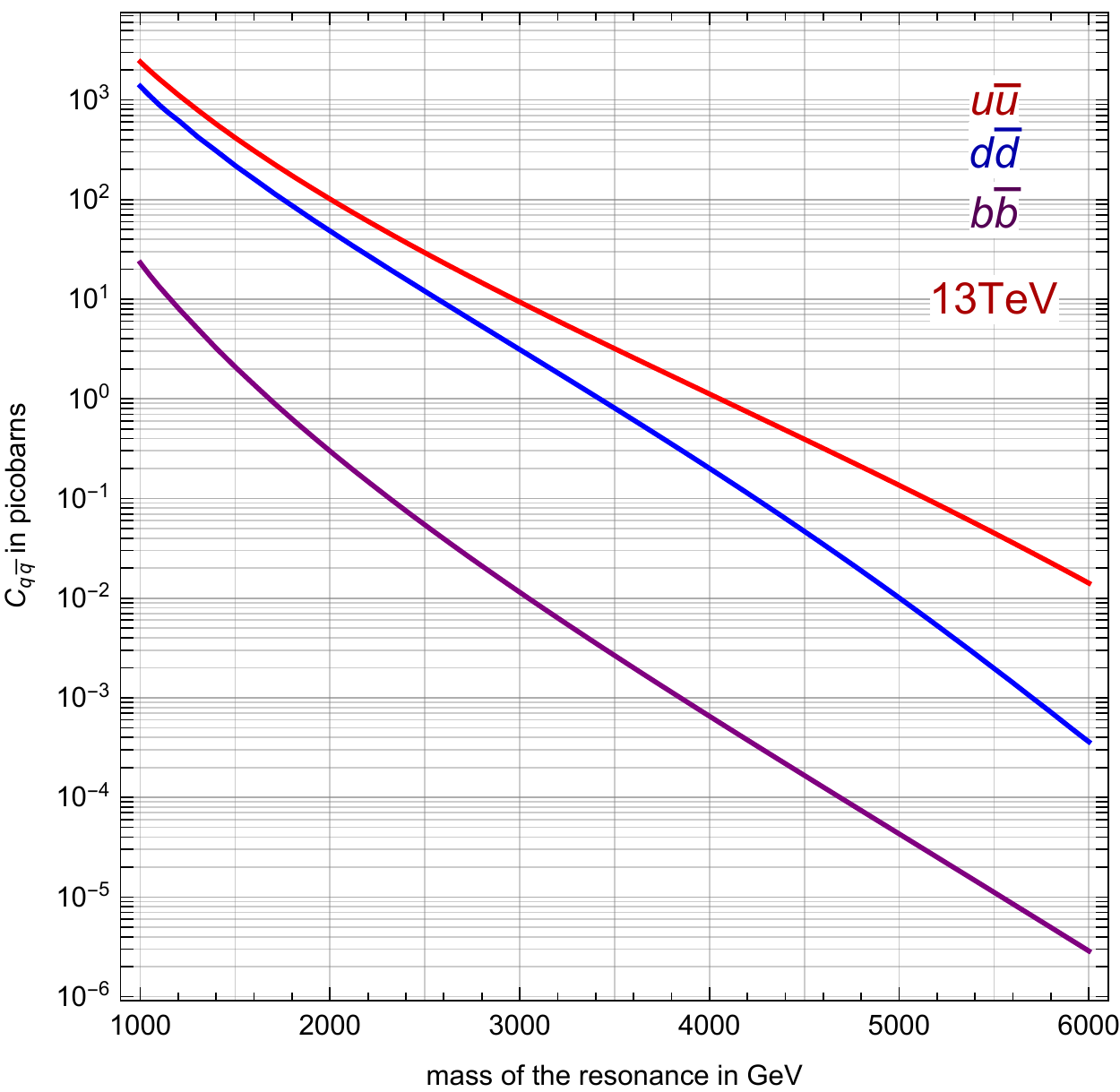}~~
\includegraphics[width=0.45\textwidth]{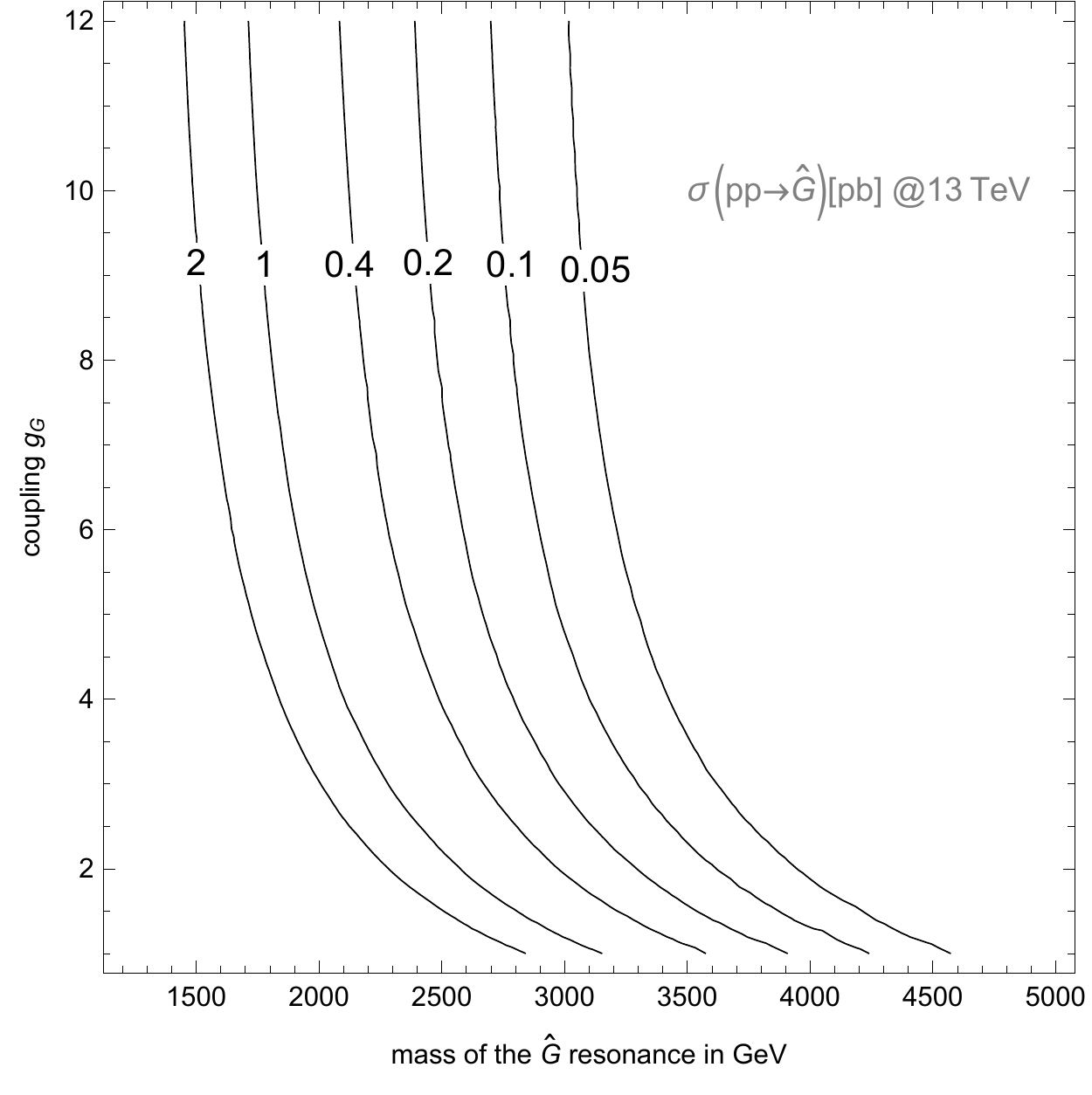}
\caption{\label{rates} Left: $C_{q\bar q}$ for different initial states at 13 TeV as function of the mass of the resonance. Right: production cross section of the $\hat G$ resonance. The coupling $\hat{g}_Gs_{q3}^2$, which controls the $b\bar b$ production channel mostly effective at large $g_G$, has been fixed to require consistency with the fit (see eq. (\ref{preferred}) with $s_{l3}=s_{q3}$.)}
\end{figure}

As mentioned in the Introduction the heavy vectors that mediate all these  low energy effective interactions should appear in direct searches at LHC. In order to compare with the collider limits on new resonances from the LHC, we need the production cross-section times the branching ratio, since experimental limits are mostly given for this quantity. The production rate of an on-shell neutral spin-1 resonance $V$ of mass $M$ via $q\bar q$ initial state can be put in the  form
\be\label{total}
\sigma_{pp\to V} = C_V \sum_{q} \frac{\Gamma_{V\to q\bar{q}}}{M}\, C_{q\bar{q}}(M^2, s) 
\ee
where $C_V$ is a color factor equal to 1 for a colourless resonance and to 8 for the heavy gluon. $C_{q\bar{q}}(M^2,s)$ encodes the effects of the parton luminosities and it is given by
\be
C_{q\bar{q}}(\hat{s}, s) =\frac{4\pi^2}{3s} \int_{\hat{s}/s}^1 \frac{dx}{x} \big[ f_q(\frac{\hat{s}x}{s}) f_{\bar{q}}(x) + f_{\bar q}(\frac{\hat{s}x}{s}) f_{q}(x)\big].
\ee
$C_{q\bar{q}}$ can be computed numerically knowing the parton distribution functions. To this end  we have used the MSTW2008 PDF set \cite{MSTW} and cross-checked the numerical results with \textsc{MadGraph} \cite{madgraph}. The relevant quantities are shown in Fig.  \ref{rates}.

\medskip

According to the global symmetries of our phenomenological Lagrangian, we expect three categories of resonances, the ones of SU(4), SO(4) and U(1)$_X$. Although the gauging of the SM symmetry induces a splitting among the resonances, the phenomenology mainly depends on the $G$ representations as shown in Section~\ref{def}. All the resonances can be singly produced, except for the leptoquark and the charged vectors of SU(2)$_R$. Despite their number, the study of the present constraints can be done in a systematic way. Indeed, the properties of the resonances are set by the global symmetries and there are a few other simplifications. First, notice that the partial decay widths are linear in the mass of the resonance, which implies that the only mass dependence of the production cross section is entirely factorized into the parton luminosities computed above.  Second, all the resonances are mainly produced from a light $q\bar{q}$ initial state (with an additional component from $b\bar b$), so that the limits and production rates can be easily compared.

An overall presentation is given in Table \ref{table-resonance} where, with a few simplifications, we summarize the main characteristic of the resonances, divided according to the broken $G$ symmetry. (For more details see Appendix \ref{widths}). It is manifest that all the resonances become wide in the large-$\hat{g}$ limit, that is needed because of the fit, and they have sizeable branching ratios into third generations fermions. Because of the large width, not all of the available resonance searches can be applied, since most of them make the narrow width approximation. However there are cases such as for $t\bar t$ and $jj$, where the limits apply also for decay widths up to $\Gamma/M=0.3$ and $\Gamma/M=0.15$ respectively. Note also that we do not include possible decays into heavy composite fermions, assuming that their mass brings them above threshold.

\begin{center}
\begin{table}
\begin{center}
\begin{tabular}{c||c||c|c||c|c||c}
 & $\Gamma/M$ &BR$_{qq}$& BR$_{bb,tt}$ & BR$_{ll}$ &  BR$_{\tau\tau,\rm inv}$ & BR$_{VV}$\\ \hline
\hline

$\hat{G}$ & $ \frac{\hat{g}_{G}^2s_q^4}{24\pi}$   & $\frac{4 g_s^4}{g_G^2 g_{\hat G}^2s^2}$&$\frac{1}{2}$ &  & &  \\

\hline

$\hat{B}$ & $\frac{\hat g_G^2 (3 s_l^4 +s_q^4)}{96 \pi }$ & $\frac{44 g'^4}{3 g_G^2 \hat{g}_G^2 (3s_l^2+s_q^2)}$& $\frac{s_q^4}{2(3s_l^4+s_q^4)}$ &
$\frac{10 g'^4}{ g_G^2 \hat{g}_G^2 (3s_l^2+s_q^2)}$ &$\frac{3s_l^4}{2(3s_l^4+s_q^4)}$&  \\ \hline\hline

$\hat{X}$ & $\frac{\hat g_{X}^2(3s_q^4+s_l^4)}{48\pi}$  &$\frac{22 g'^4}{3 g_X^2 \hat{g}_X^2 (3s_q^2+s_l^2)}$& $\frac{3s_q^4}{2(3s_q^4+s_l^4)}$ &
$\frac{5 g'^4}{ g_X^2 \hat{g}_X^2 (3s_q^2+s_l^2)}$ &$\frac{s_l^4}{2(3s_q^4+s_l^4)}$&  \\ \hline\hline

$\rho^L_0$ & $\frac{c_V^2 g_\rho^2 +2 \hat{g}_\rho^2 ( 3s_q^4+s_l^4)}{96 \pi}$ &  $\frac{12 g^4}{2g_\rho^2 (c_V^2 g_\rho^2 +2 \hat{g}_\rho^2 ( 3s_q^4+s_l^4))}$& $\frac{3 s_q^4}{2  \kappa}$& $\frac{2 g^4}{2g_\rho^2 \hat g_\rho^2 \kappa}$& $\frac{ s_l^4}{2  \kappa}$ & $\frac{c_V^2 g_\rho^2}{ \hat g_\rho^2 \kappa}$ 

\end{tabular}
\caption{\label{table-resonance}Relevant quantities for the phenomenology of singly-produced neutral resonances in the model. All the formulas are given at leading order in the elementary couplings. In the last row $\kappa=(c_V^2 g_\rho^2/\hat g_\rho^2 +2  ( 3s_q^4+s_l^4))$.}
\end{center}
\end{table}
\end{center}

\subsection{A preliminary comparison with recent available data}

Following the properties listed in Table \ref{table-resonance}, it is possible to discuss the experimental constraints in a way that is quite simple and relies mainly on the global symmetries of the theory. 

\paragraph{The SU(4) resonances.}
Among the \textbf{15} of SU(4), because of the small splitting in mass, the limits are dominated by the heavy-gluon, especially in $t\bar t$ \cite{ttbar}, $b\bar b$ \cite{bbar}, $jj$ \cite{dijet}. Indeed $\hat{G}$ and $\hat{B}$ have similar dominant branching ratios to third generation fermions, with the difference that $\hat{B}$ can also decay sizeably to $\tau\bar\tau$ and dileptons. As pure limits on cross sections, dilepton searches \cite{dimuon} are the most sensitive. However, given the small branching ratio and the reduced production cross section for $\hat{B}$ as compared to $\hat{G}$, the constraints on the SU(4) resonances are dominated by the heavy gluon. A recent result for di-tau resonances \cite{ditau} has the maximum sensitivity for masses of about $1.6\ \mathrm{TeV}$ where they exclude $\sigma \mathcal{B}$ up to $8-10$ fb. However, even though $\hat{B}$ has a large branching ratio into $\tau\bar\tau$, these limits are not comparable to the ones that affect $\hat{G}$. 

We find that the parameter space of the heavy gluon is affected by $t\bar t$ searches \cite{ttbar}, a result from early Run-II, where the experimental collaboration considered decay widths up to $\Gamma/M=0.3$, that is generically predicted for all the resonances of our model. Other constraints arise from comparison with the dijet spectrum \cite{dijet}. However this search has been only done for resonances with a width up to $15\%$. If extended up to larger widths, it could be used also to constrain the $b\bar b$ decay channel.

Interestingly, $\hat{G}$ generates also the effective operator $\frac{Z}{2 M_W^2} (D_\mu G^a_{\mu\nu})^2$, with $Z=(g_s^2/g_G^2)M_W^2/m_G^2$, that modifies rapidity and invariant mass distributions of dijets. The most recent constraint has been derived in \cite{Alioli:2017jdo} and can give complementary information when the mass of the resonance is too heavy or the width is so large that interference with the QCD background can modify the dijet spectrum.

\paragraph{The U(1)$_X$ resonance.}The U(1)$_X$ vector has similar rates to $\hat B$ with the proper substitutions. Analog considerations apply.
\paragraph{The SO(4) resonances.}  In the limit where the SO(4) resonances have couplings and masses similar to the SU(4) resonances, limits from common decay channels will be still dominated by the heavy gluon. The electroweak SO(4) resonances can be looked for into dibosons \cite{Pappadopulo:2014qza,Greco:2014aza}: $W_L W_L/Z_L h$ for the neutral ones and $W_L Z_L/W_L h$ for the charged ones. Present constraints \cite{dibosons} exclude cross sections of $6\ \mathrm{fb}$ for $m_\rho= 3\, \rm TeV$.

\begin{figure}
\begin{center}
\includegraphics[width=0.45\textwidth]{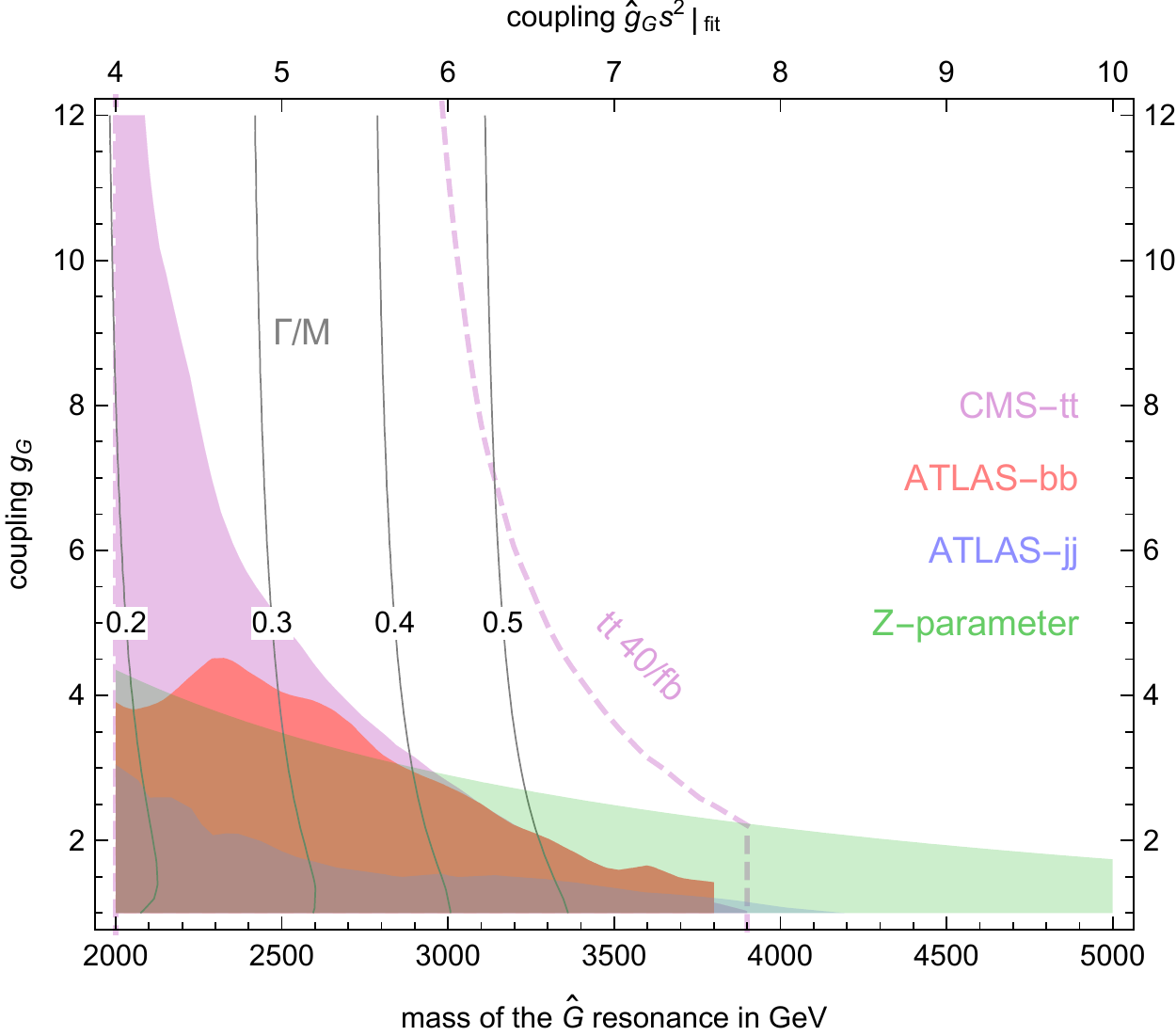}~
\includegraphics[width=0.45\textwidth]{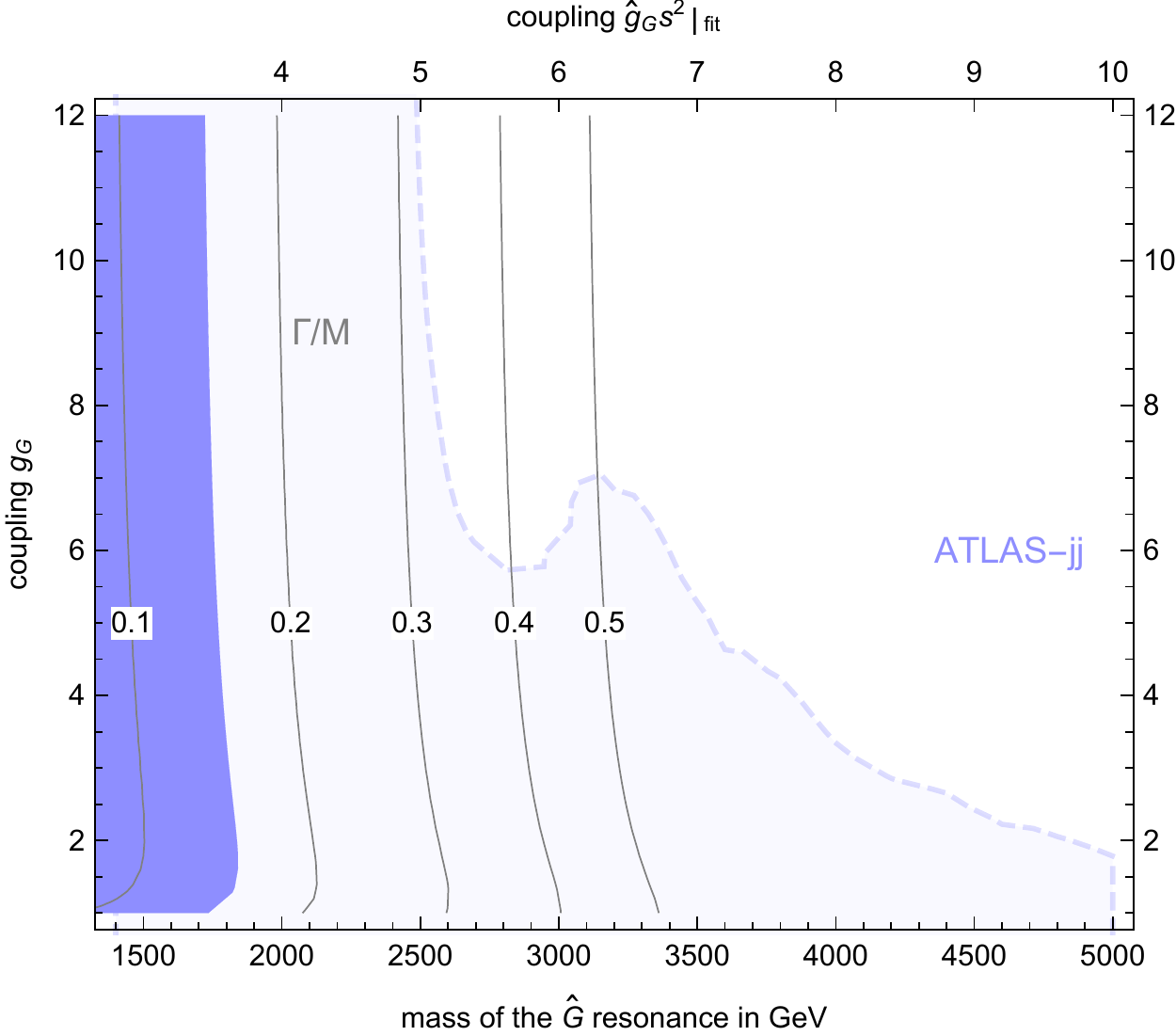}\caption{\label{LHC}Limits from LHC searches in the plane $(m_G,g_G)$. The solid lines describe $\Gamma_{\hat{G}}/m_G$ and the upper axis with $\hat{g}_G s_{q3}^2$ are set by requiring consistency with the fit (see text). \textit{Left panel}: constraints arising from searches for a resonance decaying to $t\bar t$ and $\Gamma_{\hat{G}}/m_G=0.3$ \cite{ttbar}, to $jj$ \cite{dijet} (excluding $b$-jets) and $b\bar b$ \cite{bbar} with $\Gamma_{\hat{G}}/m_G=0.15$. The limit on the $Z$ parameter uses the projection at 40 fb$^{-1}$ of \cite{Alioli:2017jdo}. \textit{Right panel}: constraints from a resonance decaying to $jj$ including the $b$-jets and $\Gamma_{\hat{G}}/m_G=0.15$ .}
\end{center}
\end{figure}

\bigskip
We show the results of a comparison with the present experimental searches in Fig.~\ref{LHC}. We present results only for the heavy gluon, since they are the dominant ones. Notice that we have also included production from $b\bar b$ initial state, which is most effective in the large $g_{G}$ limit, since it gives an irreducible contribution to the rate even in absence of a universal coupling to quarks. Such an increase contributes, for example, to set a strong exclusion bound from $t\bar t$ searches at small mass and large $g_G$ coupling.  In the left panel of figure \ref{LHC}, given the fact that $t\bar t$ searches set the strongest limits and the analysis of Ref.~\cite{ttbar} only relies on 2.6 fb$^{-1}$ of integrated luminosity, we rescale the present bounds to 40 fb$^{-1}$ to show the possible improvements in the plane $g_G,m_{\hat G}$. We used the projected bound on the $Z$ parameter at 13TeV with 40 fb$^{-1}$ from table 5 of \cite{Alioli:2017jdo}. In doing so we neglected the contribution of $b$-jets to the dijet spectrum.
In any point of the plot of figure \ref{LHC} it is possible to compute the width of the heavy gluon by requiring consistency with the fit. For example,  taking $\hat{g}_G s_q^2= 2 m_G/\mathrm{TeV}$  allows  to fix completely the predictions in terms of the two parameters of Fig.~\ref{LHC}. The right panel of Fig. \ref{LHC} gives an indication of the exclusion limits from dijet \cite{dijet} when interpreted as a constraint for the $b\bar b$ decay channel. Note however that this is meant to give only an orientation, since the search is not done in the range of $\Gamma/M$ that is needed.

\section{Summary and outlook}
\label{summary}

In this work we have offered a possible coherent description of the putative anomalies observed in charged and neutral current semi-leptonic decays of the $B$ meson, coupled to the absence of deviations from the SM in the couplings of the third generation particles to the $Z$ and the $W$. Generically we take the $B$-decay anomalies as evidence for the relevance, in flavour physics, of an approximate U(2)$^n$ symmetry. The specific key ingredient is a global Pati-Salam SU(4) symmetry under which composite vectors and fermions suitably transform. This global symmetry is supposed to emerge from a new strong dynamics, equally responsible for the existence of a composite pseudo Goldstone boson Higgs doublet.

As a way to summarise the phenomenological content of this work and to define an outlook, it is useful to group in three different sets the relevant parameters together with the observables that they mostly influence:

\begin{itemize}
\item {\it Charged current anomaly and couplings of the third generation particles to the $Z$ and the $W$.} They are controlled by the parameters
\be
(\frac{\hat{g}_G}{m_{G}},\quad \frac{\hat{g}_\rho}{m_\rho},\quad \frac{\hat{g}_X}{m_X}) s_{q3} s_{l3},\quad\quad C_{l1}=-C_{l3}
\label{par}
\ee

\item {\it Heavy gluon searches at LHC.} Other than $(\hat{g}_G/m_G) s_{q3} s_{l3}$ above, they are controlled by the parameters $g_G, m_G, s_{l3}/s_{q3}$.

\item {\it Neutral current anomaly and $\Delta C=2, \tau\rightarrow 3\mu, \tau\rightarrow \mu\gamma$.} Other than the  parameters in (\ref{par}) they involve as well the $U(2)^n$-breaking parameters $s_{l2}, s_{q2}$ and $E_{\mu 3}$.
\end{itemize}

As shown in Section \ref{delta_g_WZ}, the fit of $R_D^{(*)}$ and of $g^Z_{\tau_L}, g^Z_{\nu_\tau}, g^W_{\tau_L}$ does not fix all the parameters in (\ref{par}) but selects as preferred value $\hat{g}_G s_{q3} s_{l3}$ in a limited range around $2 m_G/{\rm TeV}$. A key element for the success of this fit, as illustrated in Fig.~\ref{FIT}, is the presence in the composite model of appropriate tree level corrections to the couplings of the third generation particles to the $Z$ and the $W$, included in the operators $\mathcal{O}^3_{Hl}, \mathcal{O}^1_{Hl}$.

The reference value $\hat{g}_G s_{q3} s_{l3}= 2 m_{G}/{\rm TeV}$ and the choice $s_{l3}/s_{q3}=1$ allow to give a preliminary discussion of what appears to be the most powerful search at LHC: the search for the heavy gluon $\hat{G}_\mu$ in terms of $g_G$ and $m_G$. The most relevant features of this search are the dominant decays  of the gluon resonance into $t \bar{t}$ and $b \bar{b}$ and its broadness. 
Because of the large width, not all of the available resonance searches can be applied. A representation of the current situation is attempted in Fig. \ref{LHC}, which must be seen as indicative of the potential of further studies, that are beyond the scope of the present paper.

The inclusion of the $U(2)^n$-breaking parameters $s_{l2}, s_{q2}$ and $E_{\mu 3}$ is crucial to get the anomaly in $R_{K^{(*)}}^{\mu e}$ and to discuss the compatibility with other flavour observables. Among them $\Delta C=2, \tau\rightarrow 3\mu, \tau\rightarrow \mu\gamma$ emerge as central to this program. The constraint  (\ref{RK}) from $R_{K^{(*)}}^{\mu e}$, which could vary by a factor of 2 depending on $\hat{g}_G/m_G$ and $\hat{g}_\rho/m_\rho$, and the bounds (\ref{boundDeltaC=2}, \ref{boundtau3mu}, \ref{boundtaumugamma}) from $\Delta C=2, \tau\rightarrow 3\mu, \tau\rightarrow \mu\gamma$ respectively appear to be all closely satisfied by 
\be
s_{l3}\approx s_{q3},\quad s_{l2}/s_{l3}\approx s_{l2}/s_{l3}\approx 0.04, \quad\quad E_{\mu 3} \approx 0.1
\ee
Any improvement of the current sensitivity on $\Delta C=2, \tau\rightarrow 3\mu, \tau\rightarrow \mu\gamma$ is likely to produce a visible signal.

On a broader prospective, assuming a strengthening  of the experimental evidence for the anomalies, an interesting theoretical question emerges. The overall description of the anomalies presented in this work is far from being possibly considered as ``UV  complete". In this respect the case is  not dissimilar from the one of composite Higgs models, often based on a global SU(3)$\times$SO(4)$\times$U(1)$_X$ symmetry group. Common to the two cases is in fact the possibility to address at the same time the ``little hierarchy problem". If this issue  is put aside, however, the problem of trying to reproduce the $B$-decay anomalies with an elementary SU(4) gauge symmetry is motivated and is actually receiving attention~\cite{Diaz:2017lit,DiLuzio:2017vat,Assad:2017iib,Calibbi:2017qbu,Bordone:2017bld} as a way to try to produce a truly UV complete description. Meanwhile one could focus on making explicit the different phenomenological expectations between an elementary and a composite SU(4) symmetry picture.

{\small
\subsubsection*{Acknowledgements}
We thank  Dario Buttazzo, Roberto Contino,  Gino Isidori, Maurizio Pierini, Gigi Rolandi, David Straub, Riccardo Torre for useful comments and discussions.

}


\appendix
\section{Generation of the effective operators  $\mathcal{O}^3_{Hl}$ and $\mathcal{O}^1_{Hl}$}
\label{AA}
In order to compute the coefficients of the operators $\mathcal{O}^{(1,3)}_{Hl}$ at the scale $m_\rho$, we need first to integrate out the vector composite resonances, and in the basis of eq.~\eqref{Ltot} this corresponds to set $\rho_\mu=e_\mu$ neglecting operators suppressed by $1/m_\rho^2$ (as operators with four heavy fermions or operators contributing to the $\hat S$ parameter). Then we need to remove the elementary composite mixing of \eqref{Lflavour}, which is in full generality
\be
\bar l_L ( \lambda_q f L_R - \tilde\lambda_q \tilde H N_R) + \mathrm{h.c.},\quad \quad L_l \to s_l l_L, \quad N_L \to - \tilde \lambda_q( \tilde H^\dag l_L)/m_\chi\,.
\ee
By making the above field redefinition in the composite lagrangian, it is evident that the only terms that can generate $\mathcal{O}^{(1,3)}_{Hl}$ are given by
\be\label{calcolo-coeff}
\mathscr{L}_{\rm composite}\supset e_\mu^a J_{\mu,\pm}^a + (i c\, d_\mu^i J_{\mu,\pm}^i +h.c.) + \frac{i}{2} \bar\chi_\pm \gamma^\mu  \overleftrightarrow{D_\mu}\chi_\pm,
\ee
where, for simplicity, we have defined fermionic currents in representations of SO(4) such as $J_{\mu,\pm}^{i}=\Psi^i_\pm\gamma^\mu \chi_\pm$ (fourplet) and  $J_{\mu,\pm}^a=\Psi_\pm\gamma^\mu T^a \Psi_\pm$ (adjoint). It is convenient to identify the Higgs current operators
\be
J_\mu^H=i H^\dag \overleftrightarrow{D_\mu} H, \quad J_\mu^{H,a}=i H^\dag\sigma^a \overleftrightarrow{D_\mu} H
\ee
and we also need the approximate expression at leading order in $1/f$ of the $d$ and $e$ symbols
\be
d_\mu^i=-\frac{\sqrt{2}}{f}(D_\mu \Pi)^i+\cdots,\quad e_\mu^{a}= A_\mu^a -\frac{i}{f^2}\Pi^T T^a D_\mu \Pi+\cdots,\quad \mathrm{with}\ \Pi^T=(h_1,h_2,h_3,h_4).
\ee
Using the above formulas we can write the terms in \eqref{calcolo-coeff} as
\begin{eqnarray}
e_\mu^a J_{\mu,+}^a &=&-\frac{1}{4f^2} J_\mu^{H,a} \bar L\gamma^\mu \sigma^a L + \frac{1}{4f^2} J_\mu^{H} \bar L\gamma^\mu L +\cdots\to -\frac{s_l^2}{4f^2}\Op^3_{Hl}+\frac{s_l^2}{4f^2}\Op^1_{Hl}\,\\
i c d_\mu^i J_{\mu,+}^i +h.c&=& -\frac{i \sqrt{2}c}{f}(D_\mu \tilde H)^t \bar L\gamma^\mu N +h.c+\cdots\to - i\sqrt{2}c\, s_l\frac{\tilde\lambda_q}{ f m_\chi} (\bar{l}_L \tilde H)(\overleftrightarrow{D_\mu} \tilde H^\dag l_L) \\
 \frac{i}{2} \bar\chi \gamma^\mu  \overleftrightarrow{D_\mu}\chi &\to& \frac{i \tilde\lambda_q^2}{2m_\chi^2} (\bar{l}_L \tilde H) (\overleftrightarrow{D_\mu} \tilde H^\dag l_L) +\cdots\,,
\end{eqnarray}
where the dots stand for subleading terms in the expansion and the arrows mean that we have made the field redefinition.
By making use of the following identity,
\be
2i \tilde H_i (\overleftrightarrow{D_\mu} \tilde H^\dag)_j = -J_\mu^{H,a} \sigma^a_{ij} +J_\mu^{H}\delta_{ij}\,,
\ee
we can then sum together all the contributions and write the coefficients of $\mathcal{O}^{(3,1)}_{Hl}$ as
\be\label{formula-C13}
C_{3l}=-V^2\frac{s_l^2}{4f^2}\big(1 +\frac{\tilde\lambda_q^2 f^2}{m_\chi^2 s_l^2} - 2\sqrt{2} c\, \frac{\tilde\lambda_q f}{m_\chi s_l}\big)\,,\quad C_{1l}=V^2\frac{s_l^2}{4f^2}\big(1 +\frac{\tilde\lambda_q^2 f^2}{m_\chi^2 s_l^2} - 2\sqrt{2} c\, \frac{\tilde\lambda_q f}{m_\chi s_l}\big).
\ee
The above result is correct at leading order and neglects the mixing of $L'$ with the $l_L$ doublet, see also \cite{Grojean:2013qca}. Notice that $C_{1l}=-C_{3l}$ can have either signs. Notice also that we do not generate operators involving the Higgs current and right-handed fermionic currents, since right-handed leptons are SO(4) singlets, and that completely analog formulas apply to the quark case.


\section{Running and operator mixing}
\label{running}
When integrating out the composite sector at the scale $M$, we generate an effective lagrangian
$V^2 \mathscr{L}=+\sum_i C_i \mathcal{O}_i$, where the operators and their coefficients $C_i$ are listed in Table \ref{operators}.

By making use of the results of \cite{RGE1,RGE2}, we can write the Renormalization Group equations for the coefficients $C_{1l},C_{3l},C_{1q}$ and $C_{3q}$, which are the most constrained by electroweak precision tests, as they affect the $W/Z$ coupling to third generation fermions. They are

\begin{eqnarray}\label{RGE}\small
16\pi^2 \frac{d C_{1l}}{d\log\mu}&=& -(2 N_c y_t^2 + \frac{2}{9} N_c g_1^2) C_1  - 2 g_1^2 C_{ll} +(2N_c y_t^2+ \frac{g_1^2}{3}) C_{1l}\,,\\
16\pi^2 \frac{d C_{3l}}{d\log\mu}&=& -(-2 N_c y_t^2 + \frac23 N_c g_2^2) C_3  + \frac23 g_2^2 C_{ll} + (2N_c y_t^2- \frac{17g_2^2}{3}) C_{3l}\,,\\
16\pi^2 \frac{d C_{1q}}{d\log\mu}&=& - (- \frac{2}{3} g_1^2) C_1 +\big[(2+4N_c)y_t^2 +\frac{2}{9}g_1^2(1+2 N_c)\big] C_{1qq} +( 6y_t^2 +\frac23 g_1^2) C_{3qq}\nonumber\\
&+&(2(N_c+2) y_t^2+ \frac{g_1^2}{3}) C_{1q}+(9y_t^2) C_{3q}\,,\\
\label{rge2}
16\pi^2 \frac{d C_{3q}}{d\log\mu}&=& -(\frac{2}{3} g_2^2) C_3  -(-2y_t^2 +\frac{2}{3}g_2^2) C_{1qq} + \big[ y_t^2 (2-4N_c) +\frac23 g_2^2(2 N_c-1)\big] C_{3qq}\nonumber\\
&+&(-\frac{3}{2}y_t^2) C_{1q}+((2Nc+2)y_t^2 + g_2^2(-\frac{17}{3})) C_{3q}\,.
\end{eqnarray}
We have neglected contributions proportional to light Yukawa couplings and electro-weak symmetry breaking effects. Notice also that, in order to simplify the equations, we did not write terms  that contribute universally to $d C_i/d\log\mu$ but would manifestly cancel in the (non-universal) observables that we are going to compute.

Solving these equations by making the approximation of keeping only the leading logarithmic dependence gives the value of the coefficients at the weak scale, that is
\be
C_i(\mu) \approx C_i(M)  - \frac{F_i}{16\pi^2} \log \frac{M}{\mu},
\ee
where $F_i$ is the right-hand side of eq.s~\eqref{RGE}-\eqref{rge2} at the scale $M$.  In turn these can be used to compute non-universal distortions in the the $W$ and $Z$ couplings given in Section \ref{delta_g_WZ},
where the left-hand side is evaluated at $\mu=M_t$ and the coefficients on the right-hand side are inputs at the scale $M=2\, \mathrm{TeV}$.

\section{Decay width of the vector resonances}
\label{widths}
Here we list the decay widths, not summed over $u,d,q,l$, of the chargeless SU(4) resonances.
The widths of the heavy gluon are:
\begin{eqnarray}
\frac{\Gamma_{\hat{G}\to u \bar u}}{m_G}&=&\frac{g_s^4}{24 \pi  g_G^2}-\frac{\hat{g}_G g_s^2 s_{q3}^2}{24 \pi  g_G}+\frac{\hat{g}_G^2 s_{q3}^4}{48 \pi },\\
\frac{\Gamma_{\hat{G}\to d \bar d}}{m_G}&=&\frac{g_s^4}{24 \pi  g_G^2}-\frac{\hat{g}_G g_s^2 s_{q3}^2}{24 \pi  g_G}+\frac{\hat{g}_G^2 s_{q3}^4}{48 \pi }
\end{eqnarray}

The widths of the vector $\hat B$ are:
\begin{eqnarray}
\frac{\Gamma_{\hat B\to u \bar u}}{m_{B}}&=&\frac{17 g'^4}{288 \pi  g_G^2}-\frac{\hat{g}_G g'^2 s_{q3}^2}{48 \sqrt{6} \pi  g_G}+\frac{\hat{g}_G^2 s_{q3}^4}{192 \pi }\\
\frac{\Gamma_{\hat B\to d \bar d}}{m_{B}}&=&\frac{5 g'^4}{288 \pi  g_G^2}-\frac{\hat{g}_G g'^2 s_{q3}^2}{48 \sqrt{6} \pi  g_G}+\frac{\hat{g}_G^2 s_{q3}^4}{192 \pi }\\
\frac{\Gamma_{\hat B\to l \bar l}}{m_{B}}&=&\frac{5 g'^4}{96 \pi  g_G^2}-\frac{\hat{g}_G g'^2 s_{l3}^2}{16 \sqrt{6} \pi  g_G}+\frac{\hat{g}_G^2 s_{l3}^4}{64 \pi }\\
\frac{\Gamma_{\hat B\to inv}}{m_{B}}&=&\frac{3 g'^4}{96 \pi  g_G^2}-\frac{\hat{g}_G g'^2 s_{l3}^2}{16 \sqrt{6} \pi  g_G}+\frac{\hat{g}_G^2 s_{l3}^4}{64 \pi }
\end{eqnarray}
The widths of the other heavy vectors can also be obtained from the couplings in Section~\ref{heavyvec} and have similar expressions.

\end{document}